\documentstyle[sprocl]{article}

\def\Journal#1#2#3#4{{#1} {\bf #2}, #3 (#4)}

\def\APP{\em Astropart. Phys.}

\def\PLB{{\em Phys. Lett.}  B}

\def\PRD{{\em Phys. Rev.} D}

\def\be{\begin{equation}}
\def\ee{\end{equation}}
\def\bea{\begin{eqnarray}}
\def\eea{\end{eqnarray}}

\begin{document}
\thispagestyle{empty}
\begin{flushright}
SU-ITP-98-48
\end{flushright}
\vspace*{.5cm}
\title{INITIAL CONDITIONS IN PRE-BIG BANG\footnote{Talk given at PASCOS-98
and at CAPP-98.}}

\author{N. KALOPER}

\address{Department of Physics, Stanford University, Stanford,
\\ CA 94305-4060, USA\\E-mail: kaloper@leland.stanford.edu}




\maketitle\abstracts{This note is a summary of the work reported
in hep-th/9801073. We give a brief discussion of the fine tuning
problem in pre-big bang cosmology. We use the flatness problem
as our test case, and in addition to the exact
numerical limits on initial conditions, we highlight
the differences between pre-big bang and standard inflation.
The main difference is that in
pre-big bang the universe must be smooth and flat in an exponentially
large domain already at the beginning of the dilaton-driven inflation.
}

It has been known at least for a decade
that it is extremely hard to construct
a working mechanism for inflation in
string theory.
The moduli fields, and in particular,
the dilaton, tend to roll rapidly
in the presence of vacuum energy, and
this rolling slows down the cosmological
expansion of the particle horizon. Hence
the standard inflationary scenarios cannot
work, since the horizon and flatness problems cannot
be solved.

One hopes that this problem is only
temporary and that it will be eventually
solved with the increasing understanding
of string theory. But there is also
a possibility that a different
implementation of inflation might be possible. One
of the most interesting suggestions in
this direction has been the pre-big bang
(PBB) scenario \cite{gv}. This
scenario attempts to use the rolling dilaton
to run inflation. Due to this, there appear
solutions which inflate superexponentially
for some time in the string frame.
In the course of this expansion,
the curvature and the coupling would become
large, and the evolution
proceeds towards the curvature singularity
(and not away from it). If one would {\it assume}
the existence of a mechanism which could
saturate the growth of the curvature
and coupling, and overturn superexponential
expansion into a subluminal power
law one, one could match the solution onto
a late time Friedmann-Robertson-Walker
(FRW) one. The phase of strong curvature
has been argued to resemble the Big Bang
and the one preceding it has been
dubbed the pre-big bang \cite{gv}.

Nevertheless, it has not been clear just how successful
PBB is in explaining cosmological
problems. In fact already in \cite{gv} it has been
noted that at the onset of
the dilaton-dominated inflating phase, the size of the
homogeneous domain had to be large
in string units, $L_H >>l_s$,
and the string coupling had to be small, $g^2 <1$, but
this has not been considered a real problem. More recently,
this has been revisited in \cite{tw}, where it has been argued
that in a universe which was not always dilaton-dominated,
the requirements for sufficient inflation would require fine
tuning of initial conditions. The main point
of \cite{tw} was that the presence of matter other than the
dilaton could delay the onset of inflation and hence
infringe on the efficiency of the scenario. Counterarguments
have been given in \cite{ven,bmuv,ms}, where it has been
claimed that there exist initial conditions for which
this does not happen.

We have undertaken a comprehensive analysis of these issues
in \cite{klb}, and have found a precise quantitative formulation
of the fine tuning problem in PBB. Our results show that if PBB
is to solve the horizon and flatness problems,
at the onset of the dilaton domination, the horizon size
of the inflating domain must be $L_H \ge 10^{19} l_p$. Simultaneously,
the string coupling must satisfy $0 < g^2 \le 10^{-53}$. Perhaps
the clearest way to see the severity of these constraints
is to consider the kinetic energy of the dilaton at the
onset of inflation.
In terms of the integral of motion $B = - \dot \Phi a^3$,
to be defined precisely below, this requires
$B \ge 10^{38} g^{-2} \ge 10^{91}$. These numbers are
just a reformulation, and not a solution, of the cosmological
problems. The task of PBB must be
to explain the presence of these numbers.

Let us now give a brief derivation of these constraints.
The classical dynamics of PBB is defined by the action
\be
\label{act}
S = \int d^4x \sqrt{g} \frac{e^{-\sigma}}{l^2_s} \Bigl(\frac12 R
+ \frac12 (\nabla \sigma)^2 + {\cal L}_m({\cal Y}, g_{\mu\nu},
\sigma) \Bigr)
\ee
where $g_{\mu\nu}$ and $\sigma$ are the string-frame metric and the
dilaton,
and ${\cal Y}$ stands for all other matter degrees of freedom.
This action gives a valid description of the dynamics only up to
the regions where $\exp(-\sigma) \sim 1$ and/or $R \sim 1/l^2_s$,
where higher order corrections in coupling and curvature are expected
to be important, invalidating the truncation of the theory that
led to (\ref{act}). In an FRW background
\be
\label{frw}
ds^2 = -dt^2 + a^2(t) \Bigl( \frac{dr^2}{1-kr^2} + r^2 d\Omega\Bigr)
\ee
one can write the equations of motion coming from (\ref{act}) as follows:
\bea
\label{eoms}
&&\ddot \sigma + 3H \dot \sigma
- \dot \sigma^2 = l_s^2 e^\sigma (3p-\rho) ~~~~~~~~
\dot H = H \dot \sigma - 3H^2 -
2 \frac{k}{a^2} + l_s^2 p e^\sigma \nonumber \\
&&\dot \rho + 3H(\rho + p) = 0 ~~~~~~~~~~~~
\dot \sigma^2 + 6H^2 - 6H \dot \sigma
+ 6 \frac{k}{a^2} - 2l_s^2 e^\sigma \rho = 0
\eea
If one ignores the influence of the curvature and the matter
stress-energy,
one can find two classes of expanding solutions, which are
accelerating for
$t<0$ and decelerating for $t>0$, and are separated by a curvature
singularity
at $t=0$. Strictly speaking, the solutions are valid for only for times
$|t| > l_s$. For $|t|\le l_s$, the corrections are large and
so the action (\ref{act}) does not apply. In PBB, one assumes
the existence of some corrections-driven graceful exit mechanism which
avoids the singularity, and further assumes that for $t\ge l_s \sim
l_p$ the
string coupling is roughly a constant of $O(1)$.
The period of inflation can take place between the
moment of the onset of dilaton domination, defined roughly by $\dot
\sigma^2
\sim 6k/a^2$, and at most the end of the string phase. During this time,
the evolution mostly obeys the superexponential law. If we define
$\Phi = \exp(-\sigma)/l_s^2$, we can write
\be
\label{plusb}
a_+ = |t_0| |\frac{t}{t_0}|^{1/\sqrt{3}} ~~~~~~~~ \Phi = \frac1{l_p(0)^2}
|\frac{t}{t_0}|^{\sqrt{3}+1}
\ee
Here $l_p(0) = 1/\sqrt{\Phi(t_0)}$ is the Planck length at the onset of
inflation,
as opposed to the Planck length $l_p \sim l_s$ at the end.

We can now consider the flatness problem as our test case. Requiring that
inflation
solves the flatness problem means that the the time of the exit,
$|\Omega - 1| \le 10^{-60}$. Since in PBB the exit is roughly at the
Planck time, we find that at this time $a_f H_f \sim 10^{30}$.
On the other hand, if we compute the particle horizon
during PBB, we find that
$L_f/a = {\rm const} \sim O(1)$. Also, by continuity $H_f \sim 1/l_s$,
and hence to solve the flatness problem we must have
$L_f \ge 10^{30} l_s \sim e^{70} l_s$. At the onset of inflation,
by the equations of motion, $L_i \sim a_i \sim H^{-1}_i \sim |t_i|$.
Using $L_f/a_f = L_i/a_i$, we obtain the constraint on $L_i$ at the
onset of inflation: $L_i = a_i/a_f L_f \ge 10^{30} a_i/a_f$, or
equivalently
\be
\label{constr}
\Bigl(\frac{\Phi_i}{\Phi_f}\Bigr)^{1/\sqrt{3}} =
\Bigl(\frac{|t_i|}{l_s}\Bigr)^{1+1/\sqrt{3}} \ge 10^{30}
\ee
This constraint can be translated into the following requirements
for the horizon, mass, entropy of the universe, and the coupling
constant $g$
at the onset of inflation:
\be
\label{constph}
L_i \ge 2 \times 10^{19} l_s ~~~~~  M_i \ge 10^{72} M_s
{}~~~~~ S \ge 2 \times 10^{38}~~~~~ g_0^2 \le 10^{-53}
\ee
If the universe at the onset of PBB inflation were
to appear accidentally, it would be extremely
unlikely to satisfy (\ref{constph}).
The probability for such an event, by standard inflationary arguments,
would be very low, $P \le \exp(-10^{38})$ at best. Further, it would
already have to be homogeneous over length scales of order of at least
$10^{19} l_s$, in contrast to the case of standard inflation, which
requires homogeneity over length scales of only few $l_s$.

 From inspecting the equations of motion (\ref{eoms}), it is easy to see
that the quantity $B = - \dot \Phi a^3$ is an integral of motion during
th PBB phase. Translated in terms of this independent parameter,
the inflationary constraint reads
\be
\label{Bconstr}
B \ge 4 \times 10^{38} g^{-2}_0 \ge 4 \times 10^{91}
\ee
Thus we see that if PBB is to successfully solve cosmological problems, it
must explain the origin of these numbers. Merely saying that these
numbers are required for inflation only amounts to restating the
cosmological problems in a different fashion, and no more.

We note that since this talk has been given two more
works have appeared, attempting to explain some of the large numbers
we have discussed here \cite{bdv,gpv}. This newest version of the PBB
scenario considers the collapse of classical dilaton waves, which
form black holes. However the black holes so formed must be extremely
massive, $M \sim 10^{72} M_S$, to fit our universe inside of them.
We should also note that one might be tempted to argue that
the application of the flatness problem to PBB might be misleading,
since after all at the onset of dilaton domination, the dilaton kinetic
energy and the curvature energy are of the same order of magnitude,
$\dot \sigma^2 \sim 6k/a^2$. This might resemble the standard inflation,
where at the onset of inflation $E({\rm vacuum}) \sim k/a^2$. However,
in open PBB this condition is put in by hand, by requiring that the
spacetime before inflation was precisely flat. In closed PBB,
this condition comes about only after requiring that the universe
was at least $\sim 10^{8} l_s$ across at the original singularity, again
an input. Thus the fact that the dilaton kinetic
energy and the curvature energy were approximately the same at the onset
of PBB does not explain the flatness problem.

One may wonder if these conditions could be produced by cosmological
evolution preceding inflation. The picture of the evolution before
the dilaton domination depends very strongly on the dominant source at
that time. If the universe before PBB inflation had been dominated
by a positive curvature term (i.e. had been closed) it would have had
to start from {\it another} space time singularity {\it before} the
stringy phase of PBB. At this, pre-stringy stringy time, it would have
had to be homogeneous over length scales $L \ge 10^{8} l_s$. This
may seem better than Big Bang without inflation, but is still
exponentially unlikely, with the probability
$P \sim \exp(-10^{24})$. If the universe had been dominated by
negative spatial curvature (i.e. had been open), it would have
corresponded to the collapse of a homogeneous distribution of
scalar kinetic energy for an infinite time, and of infinite size.
This universe would have to be dominated by the homogeneous
scalar fluctuation for nearly ever, in order to solve the cosmological
problems. In particular, if one considers inhomogeneous quantum
fluctuations
of the scalar field, with wavelengths $\lambda \sim |t|$ and
amplitudes $\delta \Phi \sim \sqrt{\Phi} |t|^{-1}/(2\pi)$, one could
see that they would dominate over the classical homogeneous
perturbations $\Delta \Phi \sim B/t^2$ for all times $|t| \ge B l_p(0)$.
Hence in the very far past such quantum fluctuations would backreact
with the background more strongly than the homogeneous classical
perturbations, and could completely destroy the large-scale homogeneity
for the success of PBB. We underline here that we are not addressing the
problem of growth of perturbations in this context. We do not
consider the small perturbations $\delta \rho \le \rho$, or the issues
of Jeans instability, which may or may not exist in PBB. Rather
we argue that there may be initial inhomogeneous quantum fluctuations
which can completely destroy the classical Minkowski background, making
it look like a collection of open and closed universes. Since
we do not have the necessary space to give an exhaustive discussion
of these
issues here, we refer the reader interested in a more detailed account
of this problem to \cite{klb,gpv}.

We want to emphasize that the
constraints $g^{-2} \ge
10^{53}$ and $B \ge 10^{38} g^{-2} \ge 10^{91}$ on the parameters of
the PBB theory follow from the requirement of flatness
of the universe, so they will
remain intact even if it is possible
to solve the homogeneity problem in the PBB
cosmology. We believe that the need to introduce such
extraordinarily large parameters represents a major challenge
to the PBB cosmology.

\section*{Acknowledgments}
The author wishes to thank R. Bousso and A. Linde for collaboration
on \cite{klb} and for useful conversations, A. Linde for very
helpful comments on the manuscript,
and T. Damour and G. Veneziano for useful conversations and
clarification of
their ideas.
This work has been supported in part by NSF Grant PHY-9219345.


\section*{References}
\begin{thebibliography}{99}

\bibitem{gv} G. Veneziano, \Journal{\PLB}{265}{287}{1991}; M.
Gasperini and
G. Veneziano, \Journal{\APP}{1}{317}{1993}.

\bibitem{tw} M.S. Turner and E. Weinberg, \Journal{\PRD}{56}{4604}{1997}.

\bibitem{ven} G. Veneziano, \Journal{\PLB}{406}{297}{1997}.

\bibitem{bmuv} A. Buonanno, K.A. Meissner, C. Ungarelli and G. Veneziano,
\Journal{\PRD}{57}{2543}{1998}.

\bibitem{ms} M. Maggiore and R. Sturani, \Journal{\PLB}{415}{335}{1997}.

\bibitem{klb} N. Kaloper, A. Linde and R. Bousso, eprint hep-th/9801073.

\bibitem{bdv} A. Buonanno, T. Damour and G. Veneziano, eprint
hep-th/9806230.

\bibitem{gpv} A. Ghosh, G. Pollifrone and G. Veneziano, eprint
hep-th/9806233.

\end{thebibliography}
\end{document}